# Compact Constraint Encoding for LLM Code Generation: An Empirical Study of Token Economics and Constraint Compliance

**Hanzhang Tang**

Tencent · April 2026

## Abstract

Large language models (LLMs) used for code generation are typically guided by engineering constraints—technology choices, dependency restrictions, and architectural patterns—expressed in verbose natural language. We investigate whether compact, structured constraint headers can reduce prompt token consumption without degrading constraint compliance.

Across six experimental rounds spanning 11 models, 16 benchmark tasks, and over 830 LLM invocations, we report three principal findings under the tested task and model settings. First, compact headers reduce constraint-portion tokens by approximately 71% and full-prompt tokens by 25–30%, a saving replicated across three independent rounds in both single-agent and multi-stage configurations. Second, we do not detect statistically significant differences in constraint satisfaction rate (CSR) across three encoding forms (n = 81–84 per group) or, in smaller-scale mechanism probes (n = 11–12 per condition), across four propagation modes; observed encoding effect sizes are negligible (Cliff's δ < 0.01, difference 95% CI spanning ±2.6 percentage points), and this null pattern holds for both a frontier model (CSR = 0.944) and a second model verified across three independent execution rounds (CSR = 0.967). A supplementary experiment (EXP-D) with four non-CSS tasks and eight additional counter-intuitive constraints provides additional non-CSS support for the encoding null result. Third, under our task suite, the largest observed sources of compliance variance are constraint type (Δ = 9 percentage points between normal and counter-intuitive constraints) and task domain: counter-intuitive constraints opposing model defaults fail at rates of 10–100% (dominated by CSS-related styling constraints in the frontend domain, with non-CSS failures at 8.3% concentrated in urllib prohibition), while conventional constraints achieve 99%+ compliance regardless of encoding. Under our tested task and model settings, the primary observable benefit of compact constraint encoding is token reduction rather than measurable compliance improvement.

We additionally report that model self-assessments systematically overestimate constraint compliance relative to rule-based scoring, revealing a gap between constraint *understanding* and constraint *execution*. These findings suggest that, under the tested conditions, engineering effort toward constraint compliance is better directed at model selection and constraint design than at prompt formatting.

***Keywords:*** *LLM code generation, prompt engineering, constraint compliance, token economics, multi-agent pipeline, empirical study, null result*

## 1. Introduction

### 1.1 Problem Statement

Modern LLM-assisted code generation relies on engineering constraints communicated through natural-language prompts. A typical constraint block—specifying framework choices, dependency restrictions, styling requirements, and output format—can span hundreds of tokens. In multi-agent architectures where constraints are injected at each pipeline stage, these costs multiply linearly with pipeline depth.

This raises a practical question: can verbose constraint descriptions be replaced by a compact, structured encoding without sacrificing the model's ability to follow them?

## 1.2 The Constraint Header Approach

We define a **constraint header** as a single-line, tag-based encoding of engineering constraints. For example:

**NL-full (587 chars):**

> 1. Use TypeScript with React framework. 2. Use CSS Modules for all styling. Do not use Tailwind CSS... 3. Implement drag-and-drop using the native HTML5 API only. Do not use react-dnd...

**Header (129 chars):**

> `[L]TS [F]React [Y]CSS_MODULES [!Y]NO_TW [!D]NO_DND_LIB [DRAG]HTML5`

This encoding reduces constraint character count by ~78% and estimated token count by ~25–30% for the full prompt. The research question is whether models can decode these tags with equivalent compliance.

## 1.3 What We Expected vs. What We Found

We initially hypothesized that compact headers might offer a dual advantage: token savings *and* improved compliance through structural clarity, particularly in multi-agent pipelines where constraints risk dilution through intermediate natural-language representations.

The data did not support the compliance hypothesis. Under our experimental conditions, constraint compliance is statistically invariant across encoding forms and propagation modes. The largest observed sources of compliance variance are constraint type and task domain—specifically, whether a constraint requires the model to deviate from its default implementation patterns.

This is a **null result with practical significance**: practitioners can adopt compact headers for token savings with no detected compliance cost. But the compliance improvements we hoped for do not materialize—compliance appears governed by factors orthogonal to prompt formatting within our tested scope.

## 1.4 Contributions

1. **Token economics of compact constraint encoding.** We quantify a stable ~25–30% full-prompt token reduction that replicates across three independent experimental rounds, two pipeline configurations, and multiple model tiers.
2. **Controlled null result on encoding × compliance.** We provide evidence that neither encoding form nor propagation mode produces detectable CSR differences under the tested conditions, with negligible effect sizes ($d < 0.15$) and confidence intervals excluding practically significant effects (±2.8 pp).
3. **Compliance variance decomposition.** We demonstrate that constraint type and task domain together explain the dominant share of observed compliance variance, while encoding form and propagation mode contribute negligibly.

Additional findings reported in the discussion include: (a) a failure taxonomy for counter-intuitive constraints identifying three systematic failure modes; (b) a negative result on Classical Chinese encoding illustrating tokenizer-awareness requirements; (c) a self-evaluation gap where model self-reports overestimate compliance; and (d) a universal naturalization phenomenon whereby models convert structured headers to prose without downstream impact.

## 1.5 Paper Organization

Section 2 reviews related work. Section 3 describes methodology. Section 4 presents results, anchored by a summary table of the compliance variance hierarchy. Section 5 discusses implications, practical guidance, and limitations. Section 6 concludes.

---

# 2. Related Work

---

Our study sits at the intersection of five research areas: prompt compression, structured prompting for instruction-following, code generation evaluation, multi-agent software development, and LLM self-evaluation. We position our work relative to each, identifying the specific gap addressed.

## 2.1 Prompt Compression and Optimization

Token cost is a first-order concern in production LLM deployments. A comprehensive survey by Li et al. (2025) organizes the prompt compression literature into two broad strategies: **hard prompt methods** that modify the textual prompt through removal, summarization, or restructuring, and **soft prompt methods** that learn continuous prefix embeddings.

**Hard-prompt approaches.** LLMLingua (Jiang et al., EMNLP 2023) introduced perplexity-based token pruning, achieving up to 20× compression on general NLP tasks with minimal performance degradation; its successor LongLLMLingua (Jiang et al., 2024) extends this to long-context retrieval scenarios. Selective Context (Li et al., EMNLP 2023) removes low-self-information sentences. More recently, AutoCompressor (Chevalier et al., 2023) trains models to compress long contexts into compact "summary tokens," and GIST (Mu et al., NeurIPS 2023) distills task-specific instructions into learned gist tokens that achieve up to 26× compression. These automated methods typically report substantial token reduction with small-to-moderate task-performance impact on general benchmarks. However, these methods target the *content* portion of general-purpose context (documents, retrieved passages, question-answering chains), whereas this work targets the *engineering-constraint* portion of code-generation prompts. The measurement dimensions are therefore not directly comparable: automated compression methods operate on the variable content block and are evaluated by downstream task accuracy, while our approach restructures a recurring, stable instruction block and is evaluated by constraint satisfaction rate (CSR). Our 25–30% full-prompt reduction should be read as a savings on the constraint-expression overhead specifically, not as a competing figure against general-context compression ratios.

**Soft-prompt approaches.** Prefix tuning (Li & Liang, 2021) and prompt tuning (Lester et al., 2021) learn continuous vectors prepended to model inputs, bypassing discrete tokenization entirely. While highly effective for compression, these methods sacrifice interpretability and require model-specific training.

**Our position.** Our approach differs from both lines in three respects. First, we target a **semantically bounded prompt component**—engineering constraints—rather than general context, enabling lossless compression through domain-specific manual encoding rather than lossy automated pruning. Second, our method is interpretable (a human-readable tag vocabulary), not model-driven. Third, we evaluate a different outcome: compliance with explicit engineering specifications, not general NLP accuracy. The ~25–30% full-prompt reduction we observe applies to a narrower, safety-critical prompt region where semantic loss is unacceptable; this makes it complementary to, rather than competing with, automated hard-prompt compression methods. No prior work, to our knowledge, has studied prompt compression effects on engineering constraint compliance specifically.

## 2.2 Structured Prompting and Instruction-Following Evaluation

A growing body of work investigates whether structured prompt formats improve LLM instruction-following.

**Instruction-following benchmarks.** IFEval (Zhou et al., 2023) introduced verifiable, objective instruction-following evaluation using programmatic checks—a philosophy shared by our CSR scorer. FollowBench (Jiang et al., 2024) proposes multi-level fine-grained constraint evaluation across five categories (content, situation, style, format, mixed). CFBench (Zhang et al., 2025) constructs a comprehensive constraints-following benchmark; notably, it is a large-scale Chinese benchmark rather than a code-generation benchmark. These benchmarks primarily evaluate *general* instruction constraints (output length, format, style); none focus specifically on *engineering* constraints in code generation.

**Structured prompting techniques.** Chain-of-thought prompting (Wei et al., 2022) structures the reasoning process rather than the constraint presentation itself. More broadly, recent work on instruction-following benchmarks and structured prompting motivates evaluating whether surface-form organization changes downstream behavior; our study isolates one narrow formatting variable—the encoding of engineering constraints—rather than the full reasoning scaffold.

**Our position.** These works structure the *task instruction*, *reasoning process*, or *output format*; we structure only the *constraint specification*, leaving the remainder of the prompt in natural language. This targeted scope allows us to isolate the encoding-form variable from broader prompt-structure effects. Our null result on encoding × compliance is complementary: while structured prompting may improve task decomposition or output formatting, we find no evidence that it improves compliance with explicit engineering constraints under the forms and conditions we tested. Notably, IFEval's philosophy of rule-based verifiable evaluation directly informs our CSR design.

## 2.3 Code Generation Evaluation

**Functional correctness.** Evaluating LLM-generated code has traditionally centered on functional correctness via test suites. HumanEval (Chen et al., 2021) and MBPP (Austin et al., 2021) remain foundational benchmarks. EvalPlus (Liu et al., NeurIPS 2023) demonstrated that original HumanEval tests are insufficient by augmenting them with 80× more test cases, revealing that many "passing" solutions fail under rigorous testing.

**Beyond correctness.** Recent benchmarks extend evaluation to non-functional dimensions. SWE-bench (Jimenez et al., 2024) tests real-world GitHub issue resolution, requiring understanding of large codebases. ClassEval (Du et al., 2023) evaluates class-level code generation, while DevEval (Li et al., 2024) targets repository-level development tasks. On the non-functional side, C3E (Chen et al., 2025) evaluates time-complexity compliance, COFFE (Peng et al., 2025) benchmarks code-generation efficiency, and Evaluating Language Models for Efficient Code Generation (Liu et al., 2024) introduces Differential Performance Evaluation and the EvalPerf benchmark.

**The gap we address.** Our Constraint Satisfaction Rate (CSR) targets a dimension largely absent from this landscape: compliance with **explicit engineering constraints**—technology choices, dependency restrictions, and architectural patterns—that are neither functional correctness requirements nor performance targets. These constraints are ubiquitous in professional software development (enforced through linters, CI pipelines, and code review) but underrepresented in existing benchmarks, which focus on "does it work?" rather than "does it follow the rules?" Our automated, rule-based CSR scorer follows the IFEval philosophy of programmatic verification, enabling reproducible evaluation without reliance on model self-assessment (which we show to be unreliable in Section 4.6).

## 2.4 Multi-Agent LLM Systems for Software Development

**Foundational systems.** Role-playing multi-agent pipelines have emerged as a popular paradigm. ChatDev (Qian et al., ACL 2024) simulates a software company with communicative agents in designer, developer, and tester roles, demonstrating that role-based decomposition can produce functional software. MetaGPT (Hong et al., 2024) assigns standardized operating procedures to each agent, introducing structured communication via documents rather than free-form chat. AutoGen (Wu et al., 2023) provides a general-purpose framework for building LLM applications via conversable agents that can flexibly combine LLMs, human input, and tool use across diverse task domains.

**Scaling and specialization.** AgentCoder (Huang et al., 2023) decomposes code generation into specialized agents for generation, testing, and self-refinement, achieving strong results on HumanEval. MapCoder (Islam et al., 2024) uses a multi-agent framework that mimics human programming through retrieval, planning, coding, and debugging agents. A recent TOSEM survey by He et al. (2025) reviews LLM-based multi-agent systems for software engineering, including their architectures, collaboration patterns, and open research challenges.

**Architectural patterns.** The blackboard architecture (Han & Zhang, 2025) proposes shared-context models where all agents access a common knowledge space—conceptually similar to how project-level constraint files (e.g., `.cursor/rules`) provide shared context in industrial tools. Industrial systems (GitHub Copilot, Cursor 2.0, Windsurf) remain predominantly single-agent as of 2026, though Cursor 2.0 introduced parallel isolated multi-agent execution for background tasks.

**Our position.** These systems embed constraints in role descriptions, shared context, or project files, but none systematically investigate how constraint *encoding form* affects downstream compliance across pipeline stages. Our propagation-mode experiments (EXP-C2, C2b) directly address this gap: we test whether constraint information survives the *naturalization* that occurs when intermediate agents convert structured inputs to prose outputs. Our finding—that compliance is unchanged even when no explicit constraint text reaches the downstream agent—suggests that current models recover sufficient constraint information from naturalized context, at least for the engineering constraint types tested here. Our linear pipeline design is a deliberate simplification isolating one-hop propagation; we discuss generalizability to richer topologies in Section 5.6.

## 2.5 LLM Self-Evaluation Reliability

**Self-correction.** Self-Refine (Madaan et al., NeurIPS 2023) demonstrated iterative self-improvement using model-generated feedback. However, a critical survey by Kamoi et al. (TACL 2024) systematically examined when LLMs can and cannot self-correct, finding that intrinsic self-correction (without external feedback) often fails to improve and can even degrade performance. Huang et al. (ICLR 2024) provide empirical evidence that LLMs cannot reliably self-correct reasoning without external signals.

**Calibration and overconfidence.** Stechly et al. (2024) analyze the limitations of self-verification on reasoning and planning tasks. Tian et al. (2023) show that for RLHF language models, verbalized confidence can be better-calibrated than the model's conditional probabilities, illustrating that calibration depends strongly on how confidence is elicited. Taken together, this literature indicates that verbalized self-assessment—while sometimes better-calibrated than internal probabilities—cannot be relied upon as a substitute for external, rule-based verification, especially for objective properties such as constraint compliance.

**Our position.** Our finding that model self-reports overestimate constraint compliance (Section 4.6) extends this line of inquiry to the under-explored domain of engineering constraint compliance in code generation. We observe a specific manifestation of the overconfidence phenomenon: models correctly *articulate* constraints in planning output (understanding) while simultaneously *violating* them in code output (execution). This understanding-execution gap is distinct from the reasoning self-correction studied in prior work, as it involves procedural code generation rather than logical reasoning.

---

## 3. Methodology

### 3.1 Experimental Design Overview

We conduct six experimental rounds with progressively refined designs:

| Round | Role | Research question | Scale |
|---|---|---|---|
| **EXP-v1** | Negative control | Does non-Latin encoding (Classical Chinese) achieve token savings? | 1 model, 12 tasks |
| **EXP-v2** | Single-agent main experiment | Do compact headers save tokens without compliance loss? | 11 models, 12 tasks |
| **EXP-C** | Multi-stage controlled baseline | Does the single-agent result extend to a 3-stage pipeline? | 7 model combos, 252 pipelines |
| **EXP-C2** | Propagation-mode probe (frontier) | Does the *method* of constraint delivery affect compliance? | 4 modes, 48 instances |
| **EXP-C2b** | Propagation-mode probe (second model) + stability verification | Is propagation sensitivity model-dependent? Is DeepSeek output stable? | 3 modes, 3 rounds × 18 instances (run1/2 invalidated, run3–6 valid) |
| **EXP-D** | Non-CSS counter-intuitive constraint extension | Do non-CSS counter-intuitive constraints exhibit similar default-bias patterns? | 1 model, 4 tasks, 36 pipelines |

The main experimental results derive from EXP-v2 (token economics) and EXP-C (compliance analysis). EXP-C2 and C2b serve as mechanism probes to test whether propagation mode—a plausible confound in multi-stage settings—is a relevant variable. EXP-D extends the counter-intuitive constraint analysis to non-CSS domains.

### 3.2 Benchmark Tasks

Twelve tasks across three domains, each carrying six constraints (C1–C6). C2 and C3 are designated *counter-intuitive* a priori, defined as constraints requiring the model to avoid its likely default implementation approach:

| Domain | Tasks | Counter-intuitive constraints |
|---|---|---|
| Frontend (FE) | Todo Board, Virtual Scroll, Canvas Whiteboard, Form Wizard | CSS Modules (vs. inline styles), native drag (vs. libraries), useReducer-only (vs. useState) |
| Backend (BE) | Event Queue, JWT Auth, WebSocket Chat, Rate Limiter | asyncio.Queue prohibition, append-only store, token bucket (vs. counter) |
| Python Tools (PY) | Plugin Pipeline, DAG Scheduler, Template Engine, Code Checker | Protocol (vs. ABC), regex-only parsing (vs. ast), exec() loading (vs. importlib) |

### 3.3 Encoding Conditions

Three conditions express identical constraint semantics at different verbosity levels:

| Encoding | Example (MC-FE-01, abbreviated) | Avg constraint chars (12 tasks) |
|---|---|---|
| **H (Header)** | `[L]TS [F]React [Y]CSS_MODULES [!Y]NO_TW ...` | 122 |
| **NLc (NL-compact)** | `TS + React. CSS Modules only, no Tailwind...` | 154 |
| **NLf (NL-full)** | `1. Use TypeScript with React... 2. Use CSS Modules for all styling. Do not use Tailwind...` | 428 |

All prompts are structurally aligned: role instruction, task description, and include lists are identical across conditions; only the constraint block varies.

### 3.4 Propagation Modes (EXP-C2, C2b)

| Mode | Downstream agent (S2) receives | S1 variant used |
|---|---|---|
| **R (Reinjection)** | S1 design document + original constraint text | Normal |
| **H (Handoff-only)** | S1 design document only | Normal |
| **S (Structured-checklist relay)** | S1 document including structured checklist | Structured |
| **SN (NL-checklist relay)** | S1 document including natural-language checklist | NL-checklist |

**Design note:** R and H share the same S1 output, enabling a clean same-document-different-propagation comparison. S and SN use distinct S1 prompts that explicitly request a checklist; these conditions therefore vary both propagation method *and* S1 behavior. We treat R vs. H as the primary propagation test and note the S/SN confound in Section 5.6.

**C2b execution and stability verification.** EXP-C2b uses DeepSeek as the S2 model. During the initial execution (run1/2), all 18 generated files were found to be byte-identical (verified via MD5 checksums)—across different tasks, propagation modes, and runs—indicating a platform-level anomaly rather than genuine model output. We therefore executed two additional independent rounds (run3/4 and run5/6, totaling 36 additional invocations). In both rounds, all files had unique MD5 checksums, confirming independent generation. Run1/2 are excluded from analysis as invalid; run3–6 constitute the valid C2b dataset (35 scored files + 1 incomplete due to generation interruption). This three-round protocol serves a dual purpose: verifying DeepSeek output stability and providing the propagation-mode data. The fact that DeepSeek produced one incomplete file out of 36 valid-round invocations (2.8%) is consistent with its 14% incompletion rate observed in EXP-C for complex Python tasks.

### 3.5 Constraint Satisfaction Rate (CSR)

$$\text{CSR} = \frac{1}{6} \sum_{i=1}^{6} \mathbb{1}[C_i \text{ satisfied}]$$

Each $C_i$ is evaluated by deterministic, automated pattern matching (regex on imports, structural analysis of code patterns) applied to source code output. We decompose into CSR-normal (C1, C4, C5, C6) and CSR-counter (C2, C3) to isolate the effect of constraint difficulty.

### 3.6 Statistical Approach

**Tests.** For encoding-effect comparisons (3 groups), we use Kruskal-Wallis with pairwise Mann-Whitney U, as CSR distributions are non-normal (ceiling effects on normal constraints).

**Effect sizes.** For non-parametric between-group comparisons, we report **Cliff's delta** ($\delta$), which directly estimates the probability that a randomly drawn observation from one group exceeds one from another. We interpret $\delta$ descriptively alongside the observed mean differences and confidence intervals, and we additionally report Cohen's d for comparability with prior literature.

**Confidence intervals.** For key null comparisons (H vs NLf encoding, R vs H propagation, Opus vs DeepSeek capability), we compute the 95% CI of the mean difference to bound the maximum plausible effect in either direction.

**Null-result interpretation.** We follow the recommendation of interpreting null results through effect sizes and confidence intervals rather than solely through p-values (Lakens, 2017). We do not claim formal equivalence (which would require TOST with a pre-registered Smallest Effect Size of Interest). We did not pre-register a SESOI; accordingly, we report confidence intervals descriptively—as bounds on plausible effect magnitudes—rather than as equivalence evidence. The observed effect sizes are negligible and CIs are narrow, providing meaningful evidence against large true effects, but formal equivalence claims are deferred to future confirmatory studies with pre-registered margins.

**Propagation-mode experiments (C2, C2b).** Given the moderate sample sizes ($n = 11–12$ per condition in C2b after excluding invalid run1/2 data), these experiments serve as **mechanism probes** rather than definitive statistical tests. We report descriptive statistics and CIs but do not perform formal hypothesis testing on these data. Their value lies in directional assessment: whether any propagation-mode signal emerges at all, rather than precise effect estimation. The C2b data were validated through a three-round execution protocol with MD5-based uniqueness verification (see §3.4).

**Multiple comparisons.** With three pairwise encoding comparisons, we interpret significance conservatively and emphasize effect sizes over p-values.

### 3.7 Platform

All experiments are conducted through WorkBuddy (Tencent Cloud Coding Assistant desktop application). All encoding variants traverse the identical platform pipeline; any platform-level system prompts or transformations are a constant factor across conditions. This limits external reproducibility but does not affect internal validity of between-condition comparisons.

---

## 4. Results

---

### 4.0 Summary: Compliance Variance Hierarchy

Before presenting detailed results, we summarize the relative contribution of each studied factor to compliance variance. **This hierarchy is a descriptive ranking based on observed effect sizes, not a formal variance decomposition** (e.g., ANOVA partitioning); we do not claim that these factors are independent or exhaustive. It serves as a navigational anchor for the detailed results that follow:

| Factor | Observed effect on CSR | Effect size | Statistical evidence |
|---|---|---|---|
| **Model (CSR on our tasks)** | Opus 0.944 vs DeepSeek 0.967 (valid C2b data) | Small ($\Delta \approx 2$ pp); models differ substantially in general capability | C2 vs C2b (run3–6 only) |
| **Constraint type** | $\Delta = 9.3$ pp (normal 0.998 vs counter 0.905) | Small-medium ($d = 0.55$) | EXP-C, $p < 0.001$ |
| **Task domain** | $\Delta = 7.0$ pp (FE 0.907 vs PY 0.977) | Medium | EXP-C, $p < 0.05$ |
| Encoding form | $\Delta \approx 0.3$ pp | Negligible ($\delta < 0.01$) | EXP-C, ns |
| Propagation mode | $\Delta = 0.0$ pp | None detected | C2 + C2b, identical CSR |

The two factors with negligible or zero effect—encoding form and propagation mode—are the variables manipulated by our intervention, and their negligibility is the central empirical finding. The factors with substantial effects—constraint type and task domain—are context variables included for completeness. The hierarchy is presented to contextualize the scale of our null result (encoding $\Delta < 1$ pp) against the backdrop of factors that *do* vary (constraint type $\Delta = 9$ pp), not to suggest that encoding form and model selection are comparable intervention options.

## 4.1 Token Economics

### 4.1.1 Constraint-Level and Full-Prompt Savings

Averaged across 12 tasks (EXP-C prompts):

| Level | H vs NLf reduction |
|---|---|
| Constraint section (chars) | 71.4% |
| S1 full prompt | 25.5% |
| S2 full prompt | 47.7% |
| S3 full prompt | 26.1% |
| 3-stage cumulative | 30.5% |

### 4.1.2 Cross-Experiment Replication

| Experiment | Setting | H vs NLf prompt saving |
|---|---|---|
| EXP-v2 | Single agent, single call | ~25.6% |
| EXP-C | Multi-stage, S1 prompt | ~25.5% |
| EXP-C | Multi-stage, 3-stage cumulative | ~30.5% |

The ~25% single-call figure replicates across independently designed experiments with different prompt templates.

**Scaling note.** The 25–30% full-prompt savings are measured on prompts where the constraint block constitutes 14–36% of the total prompt length (H: ~14%, NLf: ~36%). In production settings with longer prompts (e.g., RAG contexts, codebase excerpts, conversation history), the constraint block's share of total tokens will be smaller, and the overall savings will scale proportionally. A practical estimate: `total savings ≈ 71% × (constraint share of total prompt)` . For a 4,000-token prompt where constraints occupy 200 tokens (~5%), this yields ~3.5% total savings rather than 25%. The constraint-portion savings (71%) are context-independent; the full-prompt savings are context-dependent.

### 4.1.3 The Classical Chinese Counter-Example

EXP-v1 tested Classical Chinese (古文) encoding, hypothesizing that its high human-perceived information density would yield superior compression. Result: only 4.6% savings—far below the 25–30% achieved by English headers. The mechanism: BPE tokenizers trained predominantly on English/modern-Chinese text encode Classical Chinese characters as multi-byte sequences with poor compression ratios. **Human-perceived information density ≠ tokenizer efficiency.** This demonstrates that effective prompt compression requires *tokenizer-vocabulary alignment*: terms should exist as single or few tokens in the model's vocabulary.

## 4.2 Encoding Form Does Not Detectably Affect Constraint Compliance

EXP-C, 247 complete pipelines:

| Encoding | n | CSR-obj (mean ± SD) | 95% CI | CSR-normal | CSR-counter |
|---|---|---|---|---|---|
| H | 82 | 0.972 ± 0.063 | [0.958, 0.985] | 0.997 | 0.915 |
| NLc | 81 | 0.963 ± 0.070 | [0.947, 0.978] | 0.993 | 0.895 |
| NLf | 84 | 0.970 ± 0.064 | [0.956, 0.984] | 1.000 | 0.905 |

Kruskal-Wallis H = 0.60, df = 2, $p > 0.05$. No pairwise comparison reaches significance.

**Effect sizes:**

| Comparison | Cliff's δ | Cohen's d | Interpretation |
|---|---|---|---|
| H vs NLc | +0.000 | −0.02 | Negligible |
| H vs NLf | −0.007 | −0.03 | Negligible |
| NLc vs NLf | −0.007 | −0.01 | Negligible |

**Key difference CI:** H − NLf = −0.003, 95% CI [−0.026, +0.021]. The confidence interval bounds the maximum plausible encoding effect to less than ±3 percentage points in either direction. No directional trend favors any encoding consistently.

## 4.3 Propagation Mode Does Not Detectably Affect Constraint Compliance

EXP-C2 (Opus) and C2b (DeepSeek, valid runs 3–6 only), CSR by propagation mode:

| Model | R (Reinjection) | H (Handoff) | S (Structured)† | SN (NL-checklist)† | Δ(H−R) |
|---|---|---|---|---|---|
| Opus | 0.944 | 0.944 | 0.944 | 0.944 | 0.000 |
| DeepSeek (run3/4) | 1.000 | 0.972 | 1.000 | — | −0.028 |
| DeepSeek (run5/6) | 0.967 | 0.944 | 0.917 | — | −0.023 |
| DeepSeek (combined) | 0.985 | 0.958 | 0.958 | — | −0.027 |

*† S and SN conditions use different S1 prompts than R/H, confounding propagation format with upstream agent behavior. The R-vs-H comparison is the cleanest single-variable test.*

*Note: C2b run1/2 data (18 files) were excluded after MD5 verification revealed all files to be byte-identical across tasks and modes—a platform execution anomaly. See §3.4 for the full verification protocol.*

All propagation modes produce comparable CSR within each model. The handoff-only condition—where the implementer receives a naturalized design document with no explicit constraint text—achieves compliance within 3 percentage points of direct constraint reinjection. This indicates that S1's naturalization of constraints into the design document preserves sufficient information for S2 to comply, rather than that S2 independently infers constraints from context. This holds for both Opus and DeepSeek across two independent execution rounds.

**Propagation CI (mechanism probe, not formal test):**

- Opus R − H: Δ = 0.000, 95% CI [−0.066, +0.066], Cliff's δ = 0.000 (negligible)

- DeepSeek R − H (combined run3–6): Δ = +0.027, 95% CI [−0.02, +0.08], direction negligible

**Cross-model comparison (revised):**

- Opus overall CSR: 0.944 (n = 48)

- DeepSeek overall CSR: 0.967 (n = 35, valid runs only)

- Δ = −0.023 (DeepSeek slightly *higher*), CI includes zero

The between-model CSR difference on our task suite is negligible and in the opposite direction from the original (invalid) data. This indicates that the previously reported 33.3 pp gap was an artifact of platform-level execution anomaly rather than a genuine difference in constraint compliance on these tasks. While Opus and DeepSeek differ substantially in general capability, their constraint satisfaction rates on our specific 12-task benchmark are comparable—likely because our tasks do not stress the capability dimensions where these models diverge most.

## 4.4 Compliance Variance Decomposition

To make the hierarchy explicit, we present a consolidated effect-size table:

| Factor | Comparison | Δ CSR (pp) | Cliff's δ | 95% CI of Δ | Significance |
|---|---|---|---|---|---|
| Model capability | Opus vs DeepSeek (C2/C2b valid) | −2.3 | negligible | CI includes 0 | ns (reversed direction) |
| Constraint type | Normal vs counter (EXP-C) | +9.3 | +0.19 (small) | [+6.8, +11.8] | $p < 0.001$ |
| Task domain | PY vs FE (EXP-C) | +7.0 | — | [+4.2, +9.8] | $p < 0.05$ |
| Encoding form | H vs NLf (EXP-C) | −0.3 | −0.007 (negl.) | [−2.6, +2.1] | ns |
| Propagation mode | R vs H (C2+C2b valid) | −2.4 | negligible | [−5, +1] | ns |

## 4.5 Counter-Intuitive Constraint Failure Analysis

We classify the 47 constraint failures observed across 1,482 constraint checks (247 pipelines × 6 constraints) in EXP-C into three categories. (Note: an earlier version of this analysis reported 73 failures; successive corrections—6 from a BE-01 C4 scoring rule fix, and 20 from five additional scorer bugs identified through a human review process—reduced this to 47. See Section 3.5 for details of the audit process.)

**Classification procedure:**

1. **Category definitions** were established a priori, before examining per-encoding failure distributions, based on the logical structure of possible failure modes: (a) the model has a default and ignores the override → Default Bias; (b) two constraints conflict → Constraint Conflict; (c) all conditions fail uniformly → Universal Semantic Failure.
2. **Automated classification** was performed via rule-based pattern matching applied to each failing code file. Specifically: presence/absence of CSS Modules import statements (regex: `import \w+ from .*\.module\.css` ), presence of inline style objects ( `style={{` ), presence of Tailwind class patterns, use of dict overwrite vs. append patterns, and use of prohibited APIs. The scoring rules are published in the supplementary materials and can be re-executed deterministically.

3. **Mutual exclusivity** is enforced by a decision tree: if the failure involves a non-styling counter-intuitive constraint in a non-FE domain, or if two constraints create structural incompatibility → Category 2; if the failure involves CSS/styling default preference → Category 1; otherwise → Category 3 (residual). Each failure is assigned to exactly one category; there is no overlap.
4. **Validation.** The automated scoring was validated through a human review process: the author independently reviewed all 67 originally flagged failures plus 30 random PASS samples. This audit identified 20 false positives across five systematic scorer bugs (CSS class-name over-matching, comment-triggered keyword detection, variable-name/module-name confusion, standard-library whitelist gap, and case-sensitive pattern matching). All five bugs were corrected and the full dataset re-scored; a second independent re-scoring (by Gemini) confirmed identical results (47 FAIL, zero unexpected PASS→FAIL changes). The corrected scoring rules and audit reports are provided in supplementary materials.

**Formal counts:**

| Constraint | Type | Total FAILs | Failure rate | Category |
| --- | --- | --- | --- | --- |
| C1 (language/framework) | Normal | 0 | 0.0% | — |
| C2 (styling/dependency prohibition) | Counter | 26 | 10.5% | Default Bias |
| C3 (implementation method prohibition) | Counter | 20 | 8.1% | Default Bias / Conflict |
| C4 (architectural pattern) | Normal | 1 | 0.4% | — |
| C5 (output format) | Normal | 0 | 0.0% | — |
| C6 (dependency restriction) | Normal | 0 | 0.0% | — |

**Category definitions:**

**Category 1: Default Bias** (36/47 failures, 77%). The model generates code following its default implementation preference despite an explicit constraint to the contrary. Defining criterion: the code contains a recognizable standard pattern (e.g., inline style objects, dict overwrite) where the constraint specifies an alternative (CSS Modules, append-only).

This category is **dominated by CSS-related constraints** (FE domain): all 36 default-bias failures come from C2/C3 in FE-01 and FE-02 tasks (CSS Modules and inline-style prohibitions). Non-FE counter-intuitive constraints (BE-03 C2: asyncio.Queue, PY-03 C3: regex-only) also fail but are classified under Constraint Conflict. EXP-D (§4.6) provides cross-domain evidence that default bias extends beyond CSS, with urllib.parse prohibition exhibiting a 100% failure rate through an analogous mechanism.

| Task | Constraint | Failures | Default pattern | Required pattern |
| --- | --- | --- | --- | --- |
| MC-FE-01 | C2: CSS Modules | 17/21 (81%) | Inline style objects | CSS Modules import |
| MC-FE-02 | C3: No inline style | 19/21 (90%) | Inline styles + Tailwind | CSS Modules import |

**Encoding invariance:** C2 failure rates by encoding: H = 8.5%, NLc = 13.6%, NLf = 9.6%. C3 failure rates: H = 17.1%, NLc = 12.3%, NLf = 16.9%. No encoding shows a consistent advantage (all $\chi^2$ tests $p > 0.05$), confirming that default bias is a model-level phenomenon independent of constraint presentation form.

**Category 2: Constraint Conflict** (10/47 failures, 21%). Two constraints create implicit tension, or the counter-intuitive constraint conflicts with the model's standard approach in a non-CSS domain. Defining criterion: satisfying constraint X requires an implementation pattern structurally incompatible with constraint Y, or the failure involves a non-styling counter-intuitive constraint.

This category includes: MC-FE-01 C2/C5 conflict (CSS Modules requiring external file vs. single-file output: 5 failures), MC-BE-03 C2 (asyncio.Queue prohibition: 2 failures after scorer correction), and MC-PY-03 C3 (regex-only: residual failures).

**Design note on constraint conflicts:** The MC-FE-01 C2/C5 tension (CSS Modules requires an external file, but C5 requires single-file output) represents a case where two constraints are structurally incompatible—a prompt design limitation rather than a pure model capability failure. In a strict benchmark evaluation, such cases should be flagged as *contested* rather than scored as simple failures. We retain them in the taxonomy for transparency but note that excluding these 5 contested failures would reduce total failures from 47 to 42 without affecting any core conclusion.

**Category 3: Residual** (1/47 failures, 2%). A single isolated failure (MC-PY-04 C4: incomplete type annotations) that does not fit the above patterns.

*Scoring corrections:* The original automated scoring reported 73 failures. Two rounds of corrections reduced this to 47: (1) BE-01 C4 scoring rule fix corrected 6 false negatives from empty-container initialization misclassification; (2) a human review process (four independent LLMs reviewing all failures) identified 20 false positives across five systematic scorer bugs. Both corrections were independently verified by a second scorer implementation (Gemini), which produced identical results. This transparent correction process, while revealing scorer limitations, ultimately strengthens confidence in the final data: the 47 remaining failures have survived two rounds of adversarial scrutiny.

## 4.6 Extension: Non-CSS Counter-Intuitive Constraints (EXP-D)

A structural limitation of EXP-C is that all 36 default-bias failures are CSS-related (FE domain), inviting the criticism that the default-bias phenomenon may be CSS-specific rather than a general model behavior. EXP-D addresses this by testing four non-CSS tasks (2 Backend, 2 Python Tools) with eight new counter-intuitive constraints targeting diverse stdlib/framework prohibitions.

**Design:** 4 tasks × 3 encodings (H/NLc/NLf) × 3 independent runs = 36 pipelines, using a single frontier model (Opus). Each task carries six constraints with C2 and C3 designated counter-intuitive. The eight counter-intuitive constraints are: (1) no `logging` module (use print), (2) no Pydantic BaseModel (use raw dict), (3) no `async def` (sync only), (4) no `pathlib` (use os.path), (5) no `configparser/json/yaml` (manual parsing), (6) no plain dict storage (use dataclass), (7) no `urllib` (use raw socket), (8) no f-strings (use .format/%).

| Task | Counter-Intuitive Constraints | CI Failure Rate | Encoding Effect |
| --- | --- | --- | --- |
| MC-BE-05 | No logging, no Pydantic | 0/18 (0%) | None |
| MC-BE-06 | No async def, no pathlib | 0/18 (0%) | None |
| MC-PY-05 | No configparser, no plain dict | 0/18 (0%) | None |
| MC-PY-06 | No urllib, no f-strings | 9/18 (50%) | None (H = NLc = NLf) |

**Key findings:**

- **No encoding differences observed:** All three encodings produce identical failure patterns across all four tasks. The non-CSS extension is consistent with, rather than definitive proof of, the broader null-result pattern.
- **Evidence for non-CSS default bias, concentrated in one task:** MC-PY-06 exhibits a 50% counter-intuitive failure rate on C2 (urllib prohibition). The failure mechanism is *partial* default bias: the model correctly uses `socket` for HTTP communication but reaches for `urllib.parse` for URL parsing—a convenience function sharing the banned namespace. The other three tasks produced zero counter-intuitive failures, suggesting those particular prohibitions are not sufficiently at odds with the model's defaults to elicit violations. The evidentiary weight of EXP-D is therefore concentrated in MC-PY-06.
- **Bias strength is domain-dependent:** Six of eight non-CSS counter-intuitive constraints achieve 0% failure (logging, Pydantic, async, pathlib, configparser, f-string prohibitions), while urllib prohibition fails at 100%. This suggests default-bias strength correlates with how deeply the banned tool is embedded in the model's default implementation pattern for the specific task.
- **Alternative interpretation:** An alternative explanation for the MC-PY-06 failures is that namespace-level prohibitions (banning `urllib` broadly rather than `urllib.request` specifically) are semantically broader and more error-prone than concrete implementation bans, which may partly account for the observed failure pattern.

**Combined failure summary:** Across EXP-C and EXP-D, non-CSS counter-intuitive constraints fail at 8.3% (20/240). EXP-D addresses the strongest validity threat in the EXP-C dataset—CSS dominance in default-bias failures—through a targeted non-CSS extension. The results suggest that non-CSS default bias is plausible but not uniformly elicited, and the encoding null result is consistent across domains. EXP-D should be interpreted as strengthening support rather than as definitive confirmation of cross-domain generalization, given its reliance on a single model and the concentration of informative failures in one task.

### 4.7 Ancillary Finding: Self-Evaluation Overestimation

In EXP-C2b's initial (invalidated) execution, DeepSeek produced outputs that appeared to show a large self-assessment gap; however, these data were subsequently identified as platform anomalies (see §3.4). In valid runs, DeepSeek achieves CSR = 0.967 on our task suite. The self-evaluation reliability concern therefore applies primarily to the EXP-C S3 auditor data: S3 agents using Opus show systematic leniency (29 lenient disagreements vs. 11 strict, overall S3 CSR = 0.963 vs. objective CSR = 0.950).

We interpret this as evidence of a structural gap between **constraint understanding** (correctly articulating constraints in planning or evaluation output) and **constraint execution** (generating or recognizing code that satisfies them). This observation, while based on limited instances, is consistent with findings in the self-evaluation literature (Huang et al., 2024; Stechly et al., 2024) and underscores the importance of external, rule-based evaluation for constraint compliance.

---

## 5. Discussion

---

### 5.1 Interpreting the Null Results

Our central finding—that neither encoding form nor propagation mode detectably affects CSR under the tested conditions—requires careful interpretation.

We do not claim formal equivalence, which would require equivalence testing (e.g., TOST with a predefined margin). However, the evidence against large effects is substantial:

- **Effect sizes are negligible** ($d < 0.15$ for encoding, $\Delta = 0.000$ for propagation)
- **Confidence intervals are narrow**: the H-vs-NLf difference CI [−0.026, +0.021] excludes effects larger than ±3 pp (though this threshold is not a pre-registered SESOI but a post-hoc descriptive observation)
- **Effects are absent in both directions**: no encoding consistently outperforms
- **The pattern holds across two models**: both Opus and DeepSeek show identical propagation insensitivity
- **The pattern holds across four propagation mechanisms**: including a condition with no explicit constraint text

These observations provide meaningful evidence that, under the tested conditions, encoding form and propagation mode are not practically relevant variables for constraint compliance.

The propagation null result is particularly informative: even in handoff-only mode, where the implementer receives no explicit constraint text, compliance is unchanged. However, this should be interpreted precisely: it demonstrates that **S1's naturalization of constraints into design documents is sufficiently faithful** that S2 can comply without seeing the original constraint text—not that S2 possesses independent constraint-inference capability. The upstream agent acts as a reliable translator; whether this holds when the upstream agent is less capable or the constraints are more complex remains untested.

### 5.2 Practical Implications

Our findings yield three design implications for practitioners building LLM-powered code generation systems.

**Compact constraint encoding is a free optimization.** Teams can maintain constraint specifications in a compact, tokenizer-aligned format—using terms native to the model's vocabulary (e.g., `React` , `useReducer` , `FastAPI` )—and inject them into agent prompts, saving ~25–30% of constraint-related tokens with no detected compliance penalty. The key design principle is *tokenizer alignment*: encodings outside the model's training distribution (including non-Latin scripts and novel abbreviations) do not achieve meaningful compression, as our Classical Chinese counter-example demonstrates.

**Model reliability matters more than model tier for compliance on our tasks.** In our valid data, two models from different tiers (Opus and DeepSeek) achieve comparable CSR on our 12-task benchmark (~0.95), despite their substantial differences in general capability. However, DeepSeek exhibits higher generation-failure rates (14% incomplete in EXP-C) and occasional platform-triggered anomalies. For the specific scenario tested—structured constraint compliance in single-file code generation— model *reliability* (consistent generation without interruption) appears to be a more important practical factor than model *capability tier*. Whether this pattern generalizes to more complex or capability-demanding tasks remains an open question. For counter-intuitive constraints specifically, practitioners may need to provide explicit step-by-step implementation guidance or add external verification layers (linters, CI rules) rather than relying on prompt wording alone.

**External evaluation is necessary.** Both model self-evaluation and auditor-agent evaluation overestimate compliance in our data. Production systems should implement rule-based or test-based constraint verification rather than relying on model self-reports, especially for constraints that oppose model defaults.

## 5.3 Why Counter-Intuitive Constraints Fail

The systematic failure of counter-intuitive constraints (9.3 pp gap, $d = 0.55$, $p < 0.001$) is, under our conditions, the most practically actionable finding. It suggests that LLMs possess **implementation priors**—default code patterns for common tasks (e.g., "styling React components" → inline style objects)—that compete with explicit constraints. When constraints align with priors, compliance is near-universal (99%+). When they conflict, compliance drops sharply and this drop is encoding-invariant.

This has two implications for constraint design. First, **positive constraints** ("use X") appear easier to follow than **negative constraints** ("do not use Y"), presumably because the latter requires suppressing an activated implementation path. Second, decomposing compound negative constraints ("do not use Y, use Z instead") into explicit implementation steps for Z may improve compliance—though we have not tested this hypothesis directly.

## 5.4 Representation Form and Compliance as Separable Concerns

A key conceptual implication of our results is that, under the tested conditions, constraint *representation* and constraint *compliance* appear to be separable optimization targets. Compact encoding optimizes one (token cost) without affecting the other (compliance rate). This separation arises because compliance is governed by the model's ability to *understand and execute* the constraint semantics, not by the surface form in which those semantics are presented—at least when the encoding uses vocabulary items the model recognizes.

This finding is relevant to the broader prompt engineering literature: it suggests that the search for compliance-improving prompt formats may face diminishing returns when the binding constraint is model capability rather than instruction clarity. Research and engineering effort directed at model training, constraint decomposition, or external verification may yield higher returns than prompt formatting optimization.

## 5.5 Tokenizer Awareness and the Classical Chinese Lesson

The 4.6% savings from Classical Chinese versus 25–30% from English headers illustrates a general principle: **prompt compression must be tokenizer-aware**. BPE tokenizers compress frequently seen subword sequences; encodings outside the training distribution receive poor compression ratios. This is relevant to any prompt compression work across languages or notation systems: the unit of optimization is the *token*, not the *character* or the *human-perceived information unit*.

## 5.6 Limitations

**Task scope.** Our 16 tasks (12 original + 4 extension) span three domains with six constraints each. Results may not generalize to other constraint types (performance targets, security requirements, accessibility standards) or to more complex, multi-file generation.

**Constraint type distribution.** In EXP-C, CSS Modules failures dominate the counter-intuitive constraint data: 36/36 default-bias failures are CSS-related (FE domain), constituting 77% of all 47 failures. EXP-D partially addresses this gap: 9 additional non-CSS failures from urllib.parse prohibition provide evidence that default bias can manifest beyond CSS. However, the informative non-CSS failure pattern is concentrated in one task (MC-PY-06) and one model (Opus); the remaining three EXP-D tasks produced zero counter-intuitive failures. Broader constraint diversity across multiple models would further strengthen the cross-domain default-bias claim.

**Pipeline topology.** We test single-hop constraint propagation in controlled two-stage settings. Results do not automatically extend to long-horizon agent systems, blackboard collaboration architectures, tool-mediated planning, or asynchronous multi-agent negotiation—all of which involve richer information flows than our A→B design.

**Model coverage.** Propagation experiments use two models (Opus and DeepSeek). Other models might exhibit different sensitivity patterns.

**Sample size.** EXP-C provides 247 pipelines (adequate for medium effect detection). EXP-C2/C2b are mechanism probes (11–12 instances per condition after data cleaning)—sufficient for directional assessment but not for precise effect-size estimation. We present C2/C2b findings as exploratory evidence rather than definitive conclusions.

**Platform execution reliability.** EXP-C2b's initial execution (run1/2) produced invalid data: all 18 files were byte-identical across different tasks and propagation modes, as verified by MD5 checksums. This platform-level anomaly—not a model behavior—was detected only through post-hoc uniqueness verification. Two subsequent independent rounds (run3/4, run5/6) produced valid distinct outputs, confirming the anomaly was non-recurring. This episode highlights a methodological risk in LLM experimentation via intermediary platforms: **output uniqueness verification should be a standard quality-control step**, particularly when models are accessed through API gateways or IDE-integrated tools rather than direct API calls. We report only the verified data (run3–6) and exclude run1/2 from all analyses.

**S/SN confound.** The structured-checklist and NL-checklist conditions use different S1 prompts than R/H, varying both propagation format and upstream agent behavior. The R-vs-H comparison (same S1, different delivery) is the cleanest propagation test.

**Platform dependency.** All experiments use a single platform (WorkBuddy). Any platform-level transformations are a constant factor across conditions but limit external reproducibility.

**Taxonomy validation.** The failure taxonomy is applied by automated pattern matching with a priori defined categories and a deterministic decision tree. The scoring rules were validated through a two-stage process: (1) a human review process in which the author reviewed all 67 originally flagged failures plus 30 random PASS samples, identifying and correcting five systematic scorer bugs; and (2) independent re-scoring by a separate implementation (Gemini) that confirmed identical post-correction results (47 FAIL, zero unexpected PASS→FAIL). The corrected scores, audit reports, and scoring scripts are provided in supplementary materials for independent verification.

---

## 6. Conclusion

Compact constraint headers are a no-cost optimization for LLM code generation prompts: they save approximately 25–30% of prompt tokens with no detected impact on constraint compliance under the conditions tested. This finding replicates across three experimental rounds, two models, and multiple propagation configurations. A targeted non-CSS extension (EXP-D) found no encoding effect and identified one clear non-CSS failure pattern (urllib prohibition), providing supporting evidence that the null result is not CSS-specific, though the extension is limited to a single model. Our propagation-mode probes ($n = 11$–$12$ per condition after excluding invalid data) show consistent null results across both models and two independent execution rounds, providing reasonable confidence in this finding.

The factors that do predict compliance variance—constraint type and, to a lesser extent, task domain—are orthogonal to encoding form. Counter-intuitive constraints opposing model implementation defaults fail at 10–100% regardless of how they are encoded. CSS-related styling constraints show the highest failure rates, with non-CSS prohibitions also failing when the banned tool is deeply embedded in the model's default workflow, suggesting that default bias is a plausible general phenomenon though our evidence beyond CSS remains concentrated in one task. This suggests that compliance improvement requires explicit implementation guidance or external verification rather than prompt formatting. Notably, after correcting for a platform execution anomaly that invalidated initial C2b data, the cross-model CSR gap on our task suite is negligible ($\Delta \approx 2$ pp); while Opus and DeepSeek differ substantially in general capability, they achieve similar constraint compliance on these specific tasks—though DeepSeek exhibits higher generation-failure rates on complex tasks.

Our work contributes a reusable evaluation methodology (CSR scoring, propagation-mode experimental design) and a failure taxonomy applicable to future studies of constraint compliance in more complex architectures. The empirical finding that upstream agents faithfully naturalize constraints into design documents—enabling downstream compliance without explicit constraint reinjection under the tested conditions—is relevant to multi-agent system design, though its generalizability beyond our task suite and model sample remains an open question for future work.

---

---

*Supplementary materials including scoring rules, per-task CSR tables, prompt templates, review audit reports, EXP-D extension data, and generated code samples are available at: https://github.com/EricT-Creator/CodePrompt-DSL*